\documentclass[12pt]{article}
\usepackage{graphicx,epsf}
\usepackage{float}
\begin{document}
\topmargin -.4in
\textheight 9.2in
\textwidth 5.5in
\sloppy
\baselineskip .25in
\begin{center} 
{\bf Plasmon exchange model for superconductivity in Carbon nanotubes}\\

{\bf S. M. Bose$^\dagger$ and S. Gayen$^*$}\\
Department of Physics, Drexel University, Philadelphia, PA 19104
\end{center}
\vskip .4in

\begin{center}
{\bf Abstract}
\end{center}
\vskip 0.3in
Recent investigations of superconductivity in carbon nanotubes have shown that
a single-wall zigzag nanotube can become superconducting at around 15 K. 
Theoretical studies of superconductivity in nanotubes using the traditional
phonon exchange model, however, give a superconducting transition temperature
$T_c$ less than 1K. To explain the observed higher critical temperature we explore the possibility of the plasmon exchange mechanism for superconductivity in
nanotubes. We first calculate the effective interaction between electrons in a 
nanotube mediated by plasmon exchange and show that this interaction can 
become attractive. Using this attractive interaction in the modified Eliashberg
theory for strong coupling superconductors, we then calculate the critical 
temperature $T_c$ in a nanotube. We find that $T_c$ is sensitively dependent
on the dielectric constant of the medium, the effective mass of the electrons and the radius of the nanotube. Our theoretical results can explain the observed
$T_c$ in a nanotube.
\vskip .4in
\noindent
$^\dagger$ Corresponding author. Telephone: +1-215-895-2718. E-mail address: 
bose@drexel.edu .\\
$^*$ On leave of absence from Scottish Church College, Kolkata, India
\newpage
\section{Introduction}
Carbon nanotubes, first dicovered by Iijima in 1991 \cite{Iijima}, are a new 
form of carbon with exotic physical properties \cite{Satio}. Depending 
upon their helicity and 
chirality the electronic and transport properties of the carbon nanotubes vary 
in spectacular ways \cite{Kane, Hamada, Dresselhaus, Odom, Ivchenko}. It has 
been shown both theoretically and experimentally that nanotubes of zero 
helicity are predominantly metallic in character whereas nanotubes of nonzero
helicity are mostly semiconducting in character \cite{Hamada,Jwm, Jwg}. 
Using different methods of synthesis 
good quality  single-wall, multi-wall as well as bundles of nanotubes have now
been produced \cite{Iijima1, Thess}. The diameter of a carbon nanotube is of 
the order of nanometer and
they are upto several microns in length. The question whether a carbon nanotube
or a group of carbon nanotubes can exhibit superconductivity has been addressed
in several recent  studies. Carbon nanotubes are observed 
to pass supercurrent between superconducting leads due to proximity effect 
\cite{Yu}. Recent experiments by Tang {\it et al.} \cite{Tang} have shown
the  presence of 
superconductivity in single-wall zigzag nanotubes of radius 2.1 $A^o$ at about 
15 K. On the other hand Kochiak {\it et al.} \cite{Kociak} have reported 
superconductivity in a bundle of arm-chair 
nanotubes of radius 7 $A^o$ at about 0.55 K. There have been some theoretical 
explanations of 
superconductivity in carbon nanotubes and the origin of superconducting 
fluctuations \cite{Benedict, Byczuk}. Sedeki {\it et al.}  \cite{Sedeki} used 
momentum space renormalization group theory to study the influence of phonons 
and the Coulomb 
interaction on the superconducting response function of armchair single-wall
 nanotubes. They found that the superconducting fluctuations due to phonons 
can be easily destroyed by Coulomb repulsion.  Gonzalez \cite{Gonzalez} has 
recently pointed out that an electron-phonon mechanism of superconductivity in 
ropes of
carbon nanotubes can give a superconducting transition temperature $T_c$ less 
than 1K. It appears that the phonon exchange mechanism can not 
account for superconductivity in a 
single-wall carbon nanotube (SWNT) obsereved at finite temperature. In this 
paper we introduce the plasmon exchange mechanism for superconductivity in a metallic carbon nanotube with the expectation that a plasmon with its frequency
higher than the phonon frequency would give a higher critical temperature in a
carbon nanotube. We first calculate the effective interaction between electrons
 in a nanotube mediated by plasmon exchange and show that this interaction can 
be attractive. We then use this effective interaction in the Eliasberg 
theory \cite{Elias} of strong coupling 
superconductors as modified by McMillan \cite{McMillan} to calculate the 
superconducting
transition temperature, $T_c$. In section 2 we introduce the plasmon exchange 
model for superconductivity in a metallic  carbon
 nanotube and show the details of our calculations. This model was previously 
used by Longe and Bose \cite{Longe} to calculate the critical temperature in 
high-$T_c$ superconductors. In section 3 we present our
 results and discussions. Finally in section 4 we present our conclusions. 

\section{The model}
Since we are going to present the plasmon-exchange model of superconductivity 
in a SWNT we first review briefly the excitation of a plasmon in a metallic 
nanotube. In 
our model we consider that the length of a carbon nanotube is very 
large (several microns) compared to its radius $a$ (several angstroms). We 
assume that the electrons can move parallel to the axis of the 
tube described by the quantum number $q$ as well as around the tube 
axis described by the azimuthal quantum number $\mu$.  The dielectric function 
 $\epsilon (Q,\omega)$ of the 
nanotube is calculated in the random phase approximation (RPA) \cite{Longe1}
using

\begin{equation}
\epsilon(Q,\omega) = \epsilon + v_o(Q) \Pi (Q,\omega),
\end{equation}
where $Q = [q,\mu/a]$, q and $\mu/a$ are the components of the wave vector for 
motions parallel and azimuthal directions, respectively. The polarization 
propagator in the frequency region of plasmon excitation has been  shown to be 
\begin{equation}
 \Pi(Q,\omega) \approx -\frac{n_s Q^2}{m \omega^2},
\end{equation}
 where 
$n_s$ and $m$ are surface number density and effective mass of the electron, 
respectively. In Eq. (1) $\epsilon$ is the dielectric constant of the medium and
 $v_o(Q)$ is the bare Coulomb interaction between two electrons on a nanotube 
and is given by  
\begin{equation}
v_o(Q)=4\pi e^2 a I_\mu(aq) K_\mu(aq)
\end{equation} 
where $I_\mu(aq)$ and $K_\mu(aq)$ are modified Bessel functions, $e$ is the 
electronic charge and the azimuthal quantum 
number $\mu$ runs through all 
integral values. The plasmon frequencies are obtained from the zeros of the 
dielectric function as
\begin{equation}
[\omega_\mu(q)]^2 = \frac{4 \pi n_s e^2 a}{m \epsilon}Q^2 I_\mu(aq) K_\mu(aq)
\end{equation}
In reference 22 it has been shown that the plasmon frequency corresponding to
$\mu = 0$ is semi-acoustic in nature whereas the frequencies for $\mu \neq 0$
are optical.
Once the dielectric function of the nanotube $\epsilon(Q,\omega)$ has been 
determined by Eq. (1), we can write the effective interaction between two 
electrons on a nanotube due to plasmon exchange as

\begin{equation}
V(Q,\omega)=\frac{v_o(Q)}{\epsilon(Q,\omega)}=\frac{v_o(Q)}{\epsilon+
v_o(Q)\Pi(Q,\omega)}
\end{equation} 

Substituting for  $\Pi(Q,\omega)$ from Eq.(2)  we can 
rewrite Eq. (5) as
\begin{equation}
V(Q,\omega)=\frac{v_o(Q)}{\epsilon} + \frac{v_o^2(Q)n_sQ^2}{\epsilon^2m\omega^2
-v_o(Q)\epsilon n_sQ^2}
\end{equation}

 The first term on the right hand side of Eq. (6) is the statically 
screened Coulomb repulsion part and the second term represents the effect of 
plasmon excitation. We 
notice that the second term can become attractive and can thus lead to 
superconductivity in a nanotube.

To examine how this effective interaction can lead to superconductivity in a 
nanotube, we use the Eliashberg model \cite{Elias} of 
superconductivity in a strong-coupling superconductor. Although in its original 
form the Eliashberg model is a numerical model, many analytic approximations 
have been presented by McMillan and others \cite{McMillan, others}. In this 
paper we use the McMillan model  which gives the critical 
temperature for superconductivity in a strong coupling superconductor as
\begin{equation}
T_c=\frac{<\omega>}{1.45}exp\big[-\frac{1.04(1+\lambda)}{\lambda-\mu^*(1+0.62
\lambda)}\big]
\end{equation}
In this equation $<\omega>$ is the average value of the frequency of the  boson,
 the exchange of which is responsible for superconductivity, $\lambda$ is the 
coupling strength due to attractive part of the effective interaction  and 
$\mu^*$ is the  Coulomb repulsion parameter. It has been shown by Allen and 
Dynes \cite{Allen} that if the effective interaction between electrons in a 
superconductor can be written as
\begin{equation}
V(Q, \omega)=v_o(Q)+\frac{2\omega(Q)|M(Q)|^2}{\omega^2-\omega^2(Q)},
\end{equation}
then the above parameters can be obtained from
\begin{equation}
\lambda=\lambda(0)=N(0)<2\frac{|M(Q)|^2}{\omega(Q)}>_{FS} 
\end{equation}
and
\begin{equation}
\lambda<\omega^2> = N(0) <2|M(Q)|^2\omega(Q)>_{FS} 
\end{equation}
where $N(0)$ is the density of states of the electrons at the Fermi surface and
$<....>_{FS}$ indicates that an average of the expression is taken over the 
Fermi surface. 
Combining Eqs. (9) and (10) one can obtain $<\omega>$ from 
\begin{equation}
<\omega>=\sqrt{\frac{\lambda <\omega^2>}{\lambda}}
\end{equation}
 We can express our effective interaction (Eq. (6)) in the form of Eq. (8) if we identify
\begin{equation}
\omega^2(Q)=\frac{n_sQ^2v_o(Q)}{m\epsilon}
\end{equation}
and 
\begin{equation}
|M(Q)|^2 = \frac{1}{2}\sqrt{\frac{n_sQ^2v_o^3(Q)}{m\epsilon^3}}
\end{equation}

Also for the carbon nanotubes where the electrons have axial and azimuthal 
motions, the Fermi surface will be cylindrical and the density of states at the Fermi surface will be given by
\begin{equation}
N(0)= \frac{m}{2\pi^2a}\sum_{\mu}\frac{1}{\sqrt{k_F^2-(\frac{\mu}{a})^2}}
\end{equation}

Substituting the values of $\omega^2(Q), |M(Q)|^2$ and $N(0)$ in 
Eqs. (9) and (10) and carrying out the average over the Fermi surface
we have calculated the value of $\lambda<\omega^2>$ and $\lambda$ and then
$<\omega>$ from Eq. (12). These parameters  obviously
depend on the dielectric constant $\epsilon$, the effective mass $m$, the
surface number density $n_s$ of the electron and the radius $a$ of the nanotube.
The Coulomb repulsion parameter $\mu^*$ depends on other boson frequencies  
 and like many other investigators \cite{McMillan,others1} 
we take its numerical value to be 0.1. Substituting these values of $<\omega>$,
$\lambda$ and $\mu^*$ in the McMillan's expression [Eq. (7)] for $T_c$, we 
obtain the critical temperature as a function of the parameters $\epsilon$,
$m$, $a$ and $n_s$.  

\section{Results and discussions}
To calculate the critical temperature $T_c$ in a nanotube one needs to know the
 values of the parameters $\epsilon$, $a$, $Z=m/m_e$ ($m_e$ being the mass of 
the bare electron) and $n_s$. It turns out that in a metallic nanotube the number density of electrons $n_s$ is fixed and is independent of whether it is an
arm-chair or zigzag nanotube. Assuming that each carbon atom in such a nanotube
 contributes one electron to the conduction band, $n_s$ can be shown to be 
$3.73 \times 10^{15}/cm^2$. The values of the other parameters are not fixed 
and are known only approximately. The radii $a$ of a nanotube is known to vary 
from 0.2 to 1.0 nm. The effective dielectric constant $\epsilon$ has been 
reported to be of the order of 1.4 \cite{Egger}. The effective mass  $Z$ has 
been reported to be of the order of that in a graphite sheet which is known to 
be 0.24 in a zigzag nanotube and speculated to be one order larger in an 
arm-chair nanotube. Since
 these parameters are not known exactly, we thought it would be interesting to 
study numerically how $T_c$ varies as a function of their reasonable (measured
or speculated) values. To get a better understanding of $\epsilon$ and $Z$ 
dependence of $T_c$, in Figure 1  we present a contour plot of $T_c$ as a 
function of $\epsilon$ and $Z$ for $a=0.21$nm corresponding to a zigzag 
nanotube. The figure clearly shows that $T_c$ 
decreases with increasing $\epsilon$ and decreasing $Z$. In 
Figure 2 we have plotted $T_c$ versus $\epsilon$
for $Z$ = 0.24, 0.26, 0.28 and 0.30  for a nanotube of radius $a= 0.21$nm 
 and in Fig. 3 we have plotted $T_c$ versus
$Z$ for $\epsilon$ = 1.3, 1.35, 1.40 and 1.45 for the same nanotube.
\begin{figure}[H]
\centering
\includegraphics[width=12cm,height=10cm]{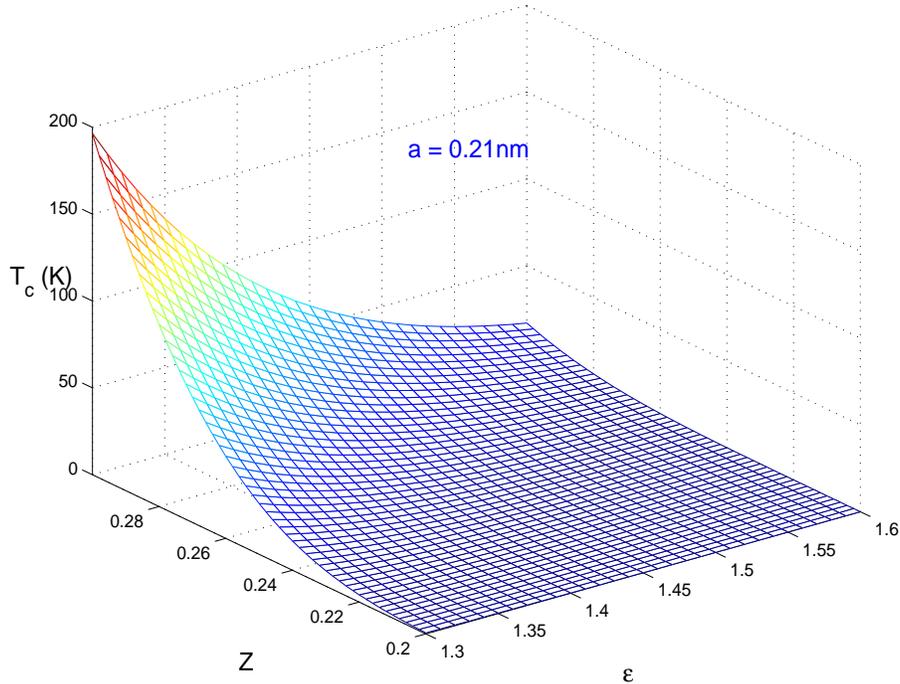}
\caption{Variation of $T_c$ as a function of $\epsilon$ and $Z$.}
\end{figure}

\begin{figure}[H]
\includegraphics[width=12cm,height=10cm]{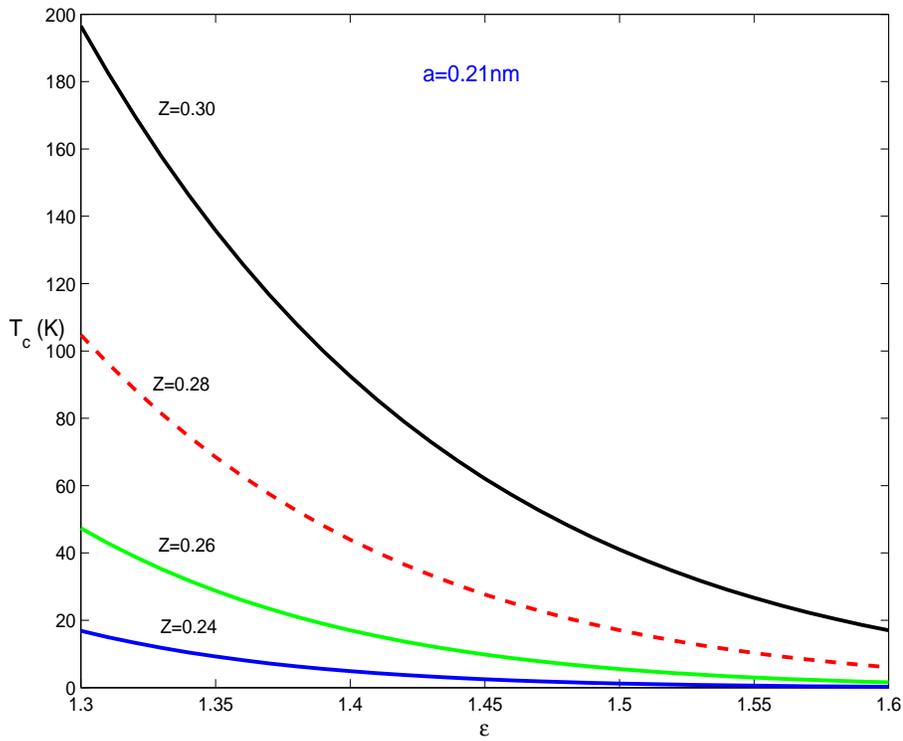}
\caption{$T_c$ as a function of $\epsilon$ for $Z=0.24, 0.26, 0.28$ and $0.30$.}
\end{figure}

\begin{figure}[H]
\includegraphics[width=13cm,height=9cm]{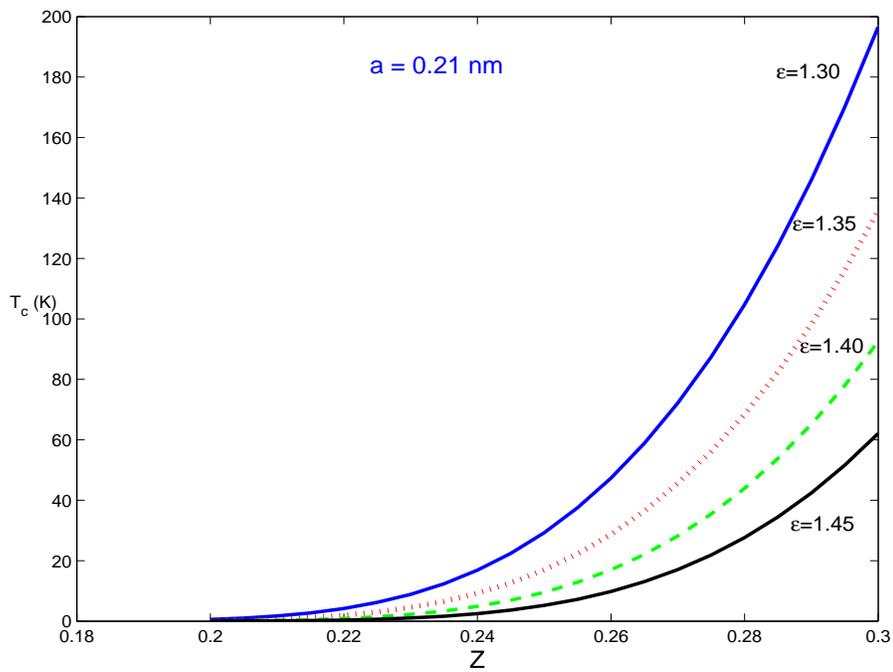}
\caption{$T_c$ as a function of $Z$ for $\epsilon =1.30, 1.35, 1.40$ and $1.45$.}
\end{figure}

 These figures make it abundantly clear that the plasmon exchange model for 
superconductivity shows that the critical temperature in a nanotube is indeed 
sensitively dependent on the parameters $\epsilon$ and $Z$ for a fixed $a$ and 
depending on their actual values $T_c$ can lie within a wide range. For example
we find that for $\epsilon$ = 1.3 and $Z$ =0.24, $T_c$ = 17 K which is close to
what Tang {\it et al.}  \cite{Tang} have measured in a single-wall zigzag 
nanotube. However, for $\epsilon$ = 1.45 and 
$Z$ =0.24, $T_c$ can be as low as 2.5 K.

\section{Summary and Conclusions}
In this paper we have studied the plasmon exchange model for superconductivity 
in a single-wall carbon nanotube. We have first shown that the effective 
interaction between two electrons mediated by plasmon exchange can become 
attractive which in its turn can lead to superconductivity in a nanotube. The 
superconducting critical temperature is then calculated by using Eliashberg 
theory for strong coupling superconductors as modified by McMillan and others. 
The critical temperature is found to be sensitively dependent on $\epsilon$, 
the dielectric constant of the medium; $m_e$, the effective mass of the 
electron; and $a$, the radius of the nanotube. For reasonable values of these 
parameters the calculated value of $T_c$ is found to be in reasonable agreement
 with the experimental values of a zigzag nanotube \cite{Tang}.


\begin{thebibliography}{99}
\bibitem{Iijima}S. Iijima, Nature  354 (1991) 56. 
\bibitem{Satio}R. Satio, G. Dresselhaus, M. S. Dresselhaus, 
{\it Physical Properties of Carbon Nanotubes}, (Imperial College Press, London, 
1999). 
\bibitem{Kane}C. Kane, L. Balents, MPA Fisher, Phys. Rev. Lett. 79 
(1997) 5086.
\bibitem{Hamada}N. Hamada, S. I. Sawada and  A. Oshiyama, Phys. Rev. Lett. 68
(1992) 1579. 
\bibitem{Dresselhaus}M. S. Dresselhaus, G. Dresselhaus, P. C. Eklund, 
{\it Science of Fullerenes and Carbon Nanotubes}, 
Academic, Sandiego, 1996.
\bibitem{Odom}T. W. Odom, J. W. Huang, P. Kim, C. M. Lieber, 
Nature 391 (1998) 62.
\bibitem{Ivchenko}E. L. Ivchenko, B. Spivak, 
Phys. Rev. B 66 (2002) 155404. 
\bibitem{Jwm}J. W. Mintmire, B. I. Dunlap, C. T. White, Phys. Rev. Lett. 
68 (1992) 631.
\bibitem{Jwg}J. W. G. Wildoer {\it et al.}, Nature 391 (1998) 59; 
T. W. Odom {\it et al.}, Nature 391 (1998) 62.
\bibitem{Iijima1}S. Iijima,T. Ichbashi, Nature 363, (1993) 603; D.S.
Bethune, C. H. Kiang, M. S. DeVries, G. Gorman, S. Savoy, R. Beyers, 
Nature 363 (1993) 605; T. W. Ebbesen, P. M. Ajayan, Nature 358 (1992) 220. 
\bibitem{Thess}A. Thess {\it et al.}, Science 273 (1996) 483; Z. F. Ren 
{\it et al.}, Science 282 (1998) 1105. 
\bibitem{Yu}A. Yu Kasumov {\it et al.}, Science 284 (1999) 1508.
\bibitem{Tang}Z. K. Tang {\it et al.}, Science 292 (2001) 2462.
\bibitem{Kociak}M. Kociak {\it et al.}, Phys. Rev. Lett. 86 (2001) 2416; 
A. Yu Kasumov {\it et al.}, Physica B 329-333 (2003) 1321.
\bibitem{Benedict}L. X. Benedict {\it et al.}, Phys. Rev. B  52 (1995) 14935.
\bibitem{Byczuk}Krzysztof Byczuk, cond-mat/ 0206086V1 (2002).
\bibitem{Sedeki}A. Sedeki, L. G. Caron, C. Bourbonnais, Phys. Rev. B  65 (2002)
140515(R).
\bibitem{Gonzalez}J. Gonzalez, Phys. Rev. Lett. 88 (2002) 076403-1.
\bibitem{Elias}G. M. Eliashberg, Zh. Eksp. Teor. Fiz. 38 (1960) 966
[Sov. Phys. JETP 11 (1960) 696];  39 (1960) 1437 [ 12 (1961) 1000].
\bibitem{McMillan}W. L. McMillan, Phys. Rev. 167 (1968) 331.
\bibitem{Longe}P. Longe, S. M. Bose, J. Phys.: Condens. Matter 4 (1992) 1811.
\bibitem{Longe1}P. Longe, S. M. Bose, Phys. Rev. B 48 (1993) 18239.
\bibitem{others}V. Z. Kresin, Phys. Lett. A 122 (1987) 434; Phys. Rev. 
B 35 (1987) 8716. 
\bibitem{Allen}P. B. Allen, R. C. Dynes, Phys. Rev. B 12 (1975) 905.
\bibitem{others1}P. Morel, P. W. Anderson, Phys. Rev. 125 63 (1962) 1263.
\bibitem{Egger}R. Egger {\it et al.}, Eur. Phys. J. B 3 (1998) 281.

\end{thebibliography}
\end{document}